\documentclass[letterpaper]{article} 
\usepackage{aaai25}  
\usepackage{times}  
\usepackage{helvet}  
\usepackage{courier}  
\usepackage[hyphens]{url}  
\usepackage{graphicx} 
\usepackage{natbib}  
\usepackage{caption} 
\usepackage{amsmath}
\usepackage{amssymb}
\usepackage{bm}
\usepackage{multirow}
\usepackage{booktabs}
\usepackage{algorithm}
\usepackage{algorithmic}
\usepackage{newfloat}
\usepackage{listings}
\DeclareCaptionStyle{ruled}{labelfont=normalfont,labelsep=colon,strut=off} 
\floatstyle{ruled}
\newfloat{listing}{tb}{lst}{}
\floatname{listing}{Listing}
\title{TechSinger: Technique Controllable Multilingual Singing Voice Synthesis \\
via Flow Matching}
\author{
    Wenxiang Guo,
    Yu Zhang,
    Changhao Pan,
    Rongjie Huang,
    Li Tang,
    Ruiqi Li,
    Zhiqing Hong,
    Yongqi Wang,
    Zhou Zhao\thanks{Corresponding author}
}
\affiliations{
    Zhejiang University \\

    \{guowx314,yuzhang34,panch,zhaozhou\}@zju.edu.cn
}

\usepackage{bibentry}

\begin{document}

\maketitle

\begin{abstract}
Singing voice synthesis has made remarkable progress in generating natural and high-quality voices. However, existing methods rarely provide precise control over vocal techniques such as intensity, mixed voice, falsetto, bubble, and breathy tones, thus limiting the expressive potential of synthetic voices. We introduce TechSinger, an advanced system for controllable singing voice synthesis that supports five languages and seven vocal techniques. TechSinger leverages a flow-matching-based generative model to produce singing voices with enhanced expressive control over various techniques. To enhance the diversity of training data, we develop a technique detection model that automatically annotates datasets with phoneme-level technique labels. Additionally, our prompt-based technique prediction model enables users to specify desired vocal attributes through natural language, offering fine-grained control over the synthesized singing. Experimental results demonstrate that TechSinger significantly enhances the expressiveness and realism of synthetic singing voices, outperforming existing methods in terms of audio quality and technique-specific control. 
\end{abstract}

\begin{links}
    \link{Code}{https://github.com/gwx314/TechSinger}
    \link{Demo}{https://gwx314.github.io/tech-singer/}
\end{links}

\section{Introduction}

Singing voice synthesis (SVS) aims to produce high-fidelity vocal performances that capture the nuances of human singing, including pitch, pronunciation, emotional expression, and vocal techniques. This field has attracted considerable attention due to its potential to revolutionize music creation and expand the boundaries of artistic expression. In recent years, rapid advancements in deep learning and generative models have driven substantial progress in singing voice synthesis \citep{resna2023multi, liu2022diffsinger, huang2022singgan, kim2023muse, hong2023unisinger}.

As singing voice synthesis technology advances, real-world applications, such as personalized virtual singers, content creation for multimedia platforms, and music production tools, highlight the growing need for controllable singing synthesis systems. However, challenges remain in achieving fine-grained control over specific vocal techniques during synthesis. Techniques like vibrato, breathy, and other stylistic nuances require precise manipulation to elevate the artistic expressiveness of synthesized singing voices. While recent algorithms have enabled accurate reproduction of acoustic features like pitch and timbre \citep{kumar2021normalization}, further advancements are needed to integrate detailed control over vocal techniques. This capability is essential for meeting the personalized and creative demands of modern music production, offering artists and creators more expressive and versatile tools for their work.

Although the task of technique-controllable singing voice synthesis holds great promise to revolutionize how we create and interact with vocal performances, it faces several significant challenges: 
1) Most existing SVS datasets, like M4Singer \citep{zhang2022m4singer} and OpenCPOP \citep{wang2022opencpop}, focus on basic features such as pitch and emotion but lack detailed annotations for singing techniques. Although Gtsinger \citep{zhang2024gtsinger} provides a dataset with several technique annotations, such datasets are still relatively rare. The absence of annotations for techniques limits models' ability to perform singing techniques.
2) Achieving fine-grained control over various singing techniques remains a core challenge. While many studies have advanced expressive singing voice synthesis by controlling features like intensity, vibrato, and breathy, they still face limitations in finely controlling multiple complex vocal techniques. Precisely modeling and reproducing various techniques while maintaining natural pitch and timbre variation is a current research focus. 
3) Utilizing the prompts for more convenient and intuitive control of singing voice synthesis based on fine-grained phoneme-level annotations is an innovative research direction \citep{wang2024prompt}. The prompt mechanism allows users to instruct the model on the desired singing style and techniques using natural language, lowering the technical barrier and enhancing user experience. However, designing effective prompt representations, training models to understand and respond to these prompts, and achieving flexible technique control while ensuring high-quality generated singing voices require further research and practice.

To address these challenges, we employ various strategies. Firstly, we tackle the scarcity of technique-annotated datasets by training a technique detector to automatically annotate technique information in open-source singing voice data. Secondly, we introduce the first flow-matching-based singing voice synthesis, enabling fine-grained control of multiple singing techniques and enhancing generated singing voices' realism and artistic expressiveness. 
To accurately model the complex relationship between pitch variations and technique expressions, we also use a flow-matching strategy to predict pitch. 
Lastly, we leverage pre-trained language models GPT-4o to construct comprehensible prompts and train a technique predictor, allowing users to easily specify desired singing styles and techniques through natural language input, thereby simplifying the operational process, enhancing user experience, and further promoting the development of personalized and customized music creation. TechSinger achieves the best results, with subjective MOS 3.89 / 4.10 in terms of the quality and technique-expressive of the singing voice generation.

In summary, this paper makes the following significant contributions to the field of singing voice synthesis:
\begin{itemize}
    \item We introduce TechSinger, the first multi-lingual singing voice synthesis model via flow matching that achieves fine-grained control over multiple techniques. 
    \item  To tackle the challenge of limited technique-annotated datasets, we develop an automatic technique detector for annotating singing techniques in open-source data.
    \item We unveil the Flow Matching Pitch Predictor (FMPP) and the Classifier-Free Guidance Flow Matching Mel-Spectrogram Postnet (CFGFMP) to improve quality.
    \item We leverage GPT-4o to create a prompt-based singing dataset and, based on this dataset, propose a technique predictor that allows for controlling singing techniques through natural language prompts.
    \item Experiments show that our model excels in generating high-quality, technique-controlled singing voices.

\end{itemize}

\section{Related Works}

\subsection{Singing Voice Synthesis}

Singing Voice Synthesis (SVS) has advanced significantly with deep learning, aiming to generate high-quality singing from musical scores and lyrics. Early models like XiaoiceSing \citep{lu2020xiaoicesing} and DeepSinger \citep{ren2020deepsinger} utilize non-autoregressive and feed-forward transformers to synthesize singing voice. VISinger \citep{zhang2022visinger} employs the VITS \citep{kim2021conditional} architecture for end-to-end SVS. GANs have also been used for high-fidelity voice synthesis \citep{wu2020adversarially, huang2022singgan}, and DiffSinger \citep{liu2022diffsinger} introduces diffusion for improved mel-spectrogram generation.
Despite these advancements, precise control over singing techniques remains a challenge, which is essential for enhancing artistic expressiveness. Controllable SVS focuses on managing aspects like timbre, emotion, style, and techniques. Existing works often target specific controls, such as Muse-SVS \citep{kim2023muse} for pitch and emotion, StyleSinger \citep{zhang2024stylesinger} and TCSinger \cite{zhang2024tcsinger} for style transfer, and models for vibrato control \citep{liu2021vibrato, song2022singing, ikemiya2014transferring}. However, we advance technique controllable SVS by enabling control over seven techniques across five languages.

\subsection{Prompt-guided Voice Generation}
In terms of voice generation, previous controls rely on texts, scores, and feature labels. Prompt-based control is emerging as a simpler, more intuitive alternative and has achieved great success in text, image, and audio generation tasks \cite{brown2020language, ramesh2021zero, kreuk2022audiogen}
In speech generation, PromptTTS \citep{guo2023prompttts} and InstructTTS \citep{yang2023instructtts} use text descriptions to guide synthesis, offering precise control over style and content.
In singing voice generation, Prompt-Singer \citep{wang2024prompt} uses natural language prompts to control attributes like the singer's gender and volume but lacks advanced technique control. This paper addresses this gap by integrating multiple techniques into prompt-based control, allowing for more sophisticated and expressive singing voice generation.

\subsection{Flow Matching Generative Models}

Flow matching \cite{lipman2022flow} is an advanced generative modeling technique that optimizes the mapping between noise distributions and data samples by ensuring a smooth transport path, reducing sampling complexity. It has significantly improved audio generation tasks.
Voicebox \citep{le2024voicebox} uses flow matching for high-quality text-to-speech synthesis, noise removal, and content editing.
Audiobox \citep{vyas2023audiobox} leverages flow matching to enhance multi-modal audio generation with better controllability and efficiency. 
Matcha-TTS \citep{mehta2024matcha} applies optimal-transport conditional flow matching for high-quality, fast, and memory-efficient text-to-speech synthesis. 
VoiceFlow \citep{guo2024voiceflow} utilizes rectified flow matching to generate superior mel-spectrograms with fewer steps.
Inspired by these successes, we use flow matching for controllable singing voice synthesis to boost quality and efficiency.

\section{Preliminary: Rectified Flow Matching}
\label{sec: pre}

Firstly, we introduce the preliminaries of the flow matching generative model \citep{liu2022flow}. When constructing a generative model, the true data distribution is $q(x_1)$ which we can sample, but whose density function is inaccessible. Suppose there is a probability path $p_t(x_t)$, where $x_0 \sim p_0(x)$ is a known simple distribution (such as a standard Gaussian distribution), and $x_1 \sim p_1(x)$ approximates the realistic data distribution. The goal of flow matching is to directly model this probability path, which can be expressed in the form of an ordinary differential equation (ODE):
\begin{equation}
\mathrm{d}x = u(x, t) \mathrm{d}t , t \in [0, 1],
\end{equation}
where $u$ represents the target vector field, and $t$ represents the time position. If the vector field $u$ is known, we can obtain the realistic data through reverse steps. We can regress the vector field $u$ using a vector field estimator $v(\cdot)$ with the flow matching objective:
\begin{equation}
    \mathcal{L}_{\mathrm{FM}}(\theta)=\mathbb{E}_{t, p_t(x)}\left\|v(x, t;\theta)-u(x, t)\right\|^2,
    \label{eq:fm_objective}
\end{equation}
where $ p_t(x)$ is the distribution of $x$ at timestep $t$. To guide the regression by incorporating a condition $c$, we can use the conditional flow matching objective \cite{lipman2022flow}:
\begin{equation}
    \mathcal{L}_{\mathrm{CFM}}(\theta)=\mathbb{E}_{t, p_1(x_1),p_t(x|x_1)}\left\|v(x, t|c;\theta)-u(x, t|x_1,c)\right\|^2,
    \label{eq:cfm_objective}
\end{equation}
Flow matching proposes using a straight path to transform from noise to data. We adopt the linear interpolation schedule between the data $x_1$ and a Gaussian noise sample $x_0$ to get the sample $x_t = (1-t)x_0 + tx_1$. Therefore, the conditional vector field is $u(x, t|x_1,c)=x_1 - x_0$, and the  rectified flow matching (RFM) loss used in gradient descent is:
\begin{equation}
    \left\|v(x, t|c;\theta)-(x_1 - x_0)\right\|^2,
    \label{eq:rfm_objective}
\end{equation}
If the vector field $u$ can be obtained, we can generate realistic data by propagating sampled Gaussian noise through various ODE solvers at discrete time steps. A common approach for the reverse flow is the Euler ODE:
\begin{equation}
x_{t+\epsilon} = x + \epsilon v(x, t | c; \theta).
\label{eq:euler}
\end{equation}
where $\epsilon$ is the step size. In this work, we use the notes, lyrics, and technique as condition $c$, while the data $x_1$ is fundamental frequencies (F0) or mel-spectrograms.

\section{TechSinger}

In this section, we outline the overall framework of TechSinger, followed by detailed descriptions of its key components, including the flow matching pitch predictor, classifier-free flow matching postnet, technique detector, and technique predictor. 
We conclude with an explanation of TechSinger's two-stage training and inference process.

\subsection{Overview}

\begin{figure*}[t]
\centering
\includegraphics[width=0.9\textwidth, trim={5mm 15mm 10mm 15mm}, clip]{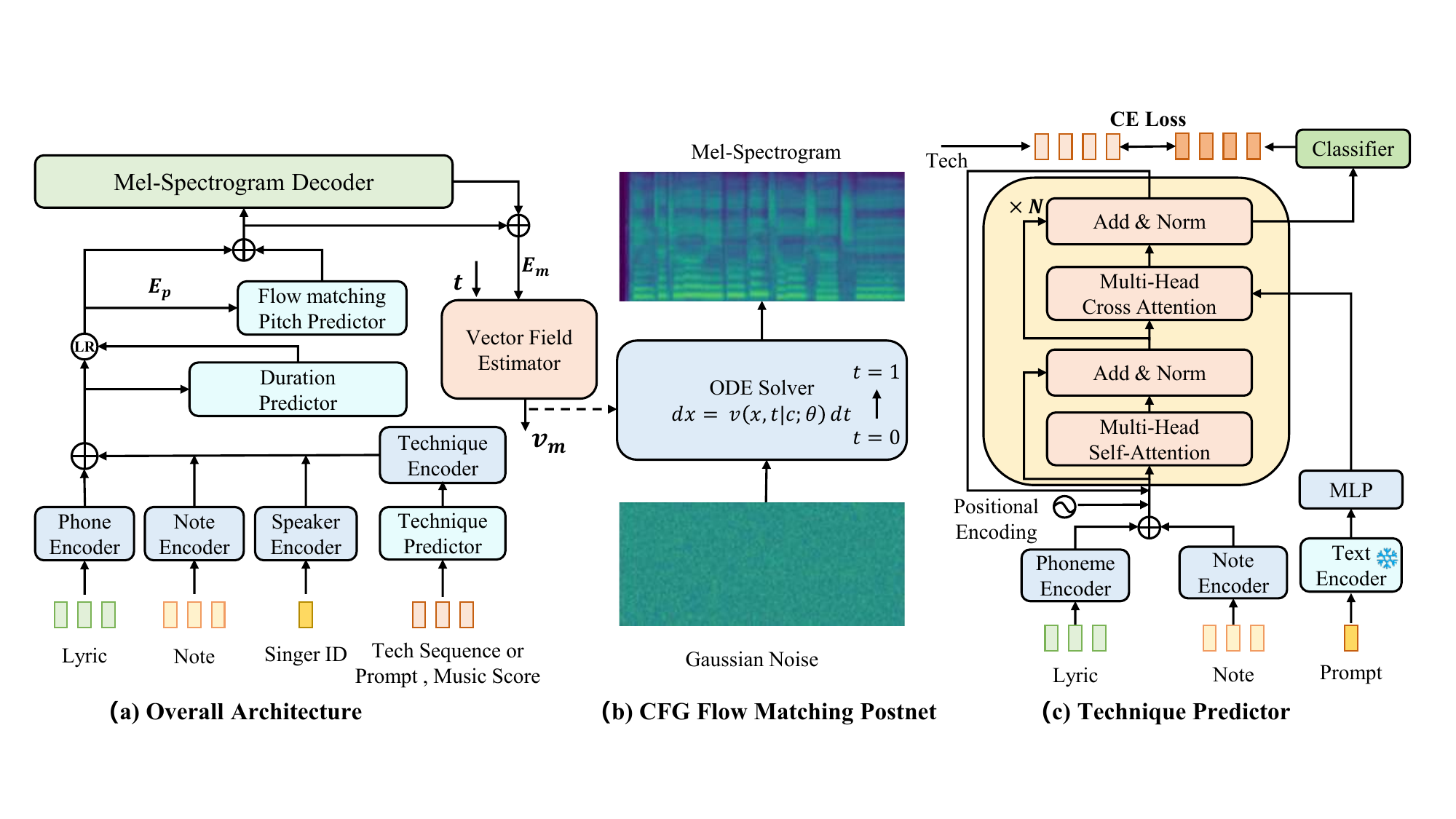}
\caption{The overall architecture of TechSinger. 
In Figure (a), the technique predictor can predict technique sequences with natural language prompts. 
The flow matching pitch predictor (FMPP) conditions on the expanded input encoding $E_p$ to generate the F0 sequences. 
The mel decoder generates the coarse mel-spectrogram.
The vector field estimator infers the vector field $v_m$.
In Figure (b), $v_m$ is used to flow the standard Gaussian noise into a fine mel-spectrogram via an ODE solver.
In Figure (c), the input of the technique predictor is prompt, note, and lyrics. 
The text encoder is a pre-trained language model.
}
\label{fig: arch}
\end{figure*}

The architecture of TechSinger is illustrated in Figure~\ref{fig: arch}. 
Initially, the phoneme encoder processes the lyrics while the note encoder captures the musical rhythm by encoding note pitches, note durations, and note types. 
Technique information is provided by encoding a sequence of techniques, and for more precise control over the singing style, a technique predictor is utilized, which generates corresponding technique sequences from the natural language prompt. 
The technique embeddings, along with the musical information, are then used to predict durations and extend to produce frame-level intermediate features $E_p$.
The flow matching-based model employs $E_p$ as the condition to generate fundamental frequencies (F0). 
Subsequently, the coarse mel decoder predicts coarse mel-spectrograms. 
Finally, the flow matching-based postnet refines these predictions to generate high-quality mel-spectrograms. 
The process concludes with the use of HiFi-GAN vocoder~\cite{kong2020hifi}, which converts the mel-spectrograms into audio signals.

\subsection{Flow Matching Pitch Predictor}
Reconstructing fundamental frequencies (F0) using only L1 loss makes it difficult to model the complex mapping between different techniques and F0. 
To precisely model the pitch contour variations across different techniques, we introduce the Flow Matching Pitch Predictor (FMPP). 
The fundamental frequency (F0) can be regarded as one-dimensional continuous data.
The corresponding condition $c$ is the combination features $E_p$ of the music score and technique sequence, and the sampled $x_1$ is the F0 extracted by open-source tool RMVPE \cite{wei2023rmvpe} as the target $f0_g$. Inspired by \citet{lipman2022flow}, we perform linear interpolation between a F0 sample $x_1 = f0_g$ and Gaussian noise $x_0$ to create a conditional probability path $x_t = (1-t)x_0 + tx_1$. We then use the vector field estimator $v_p$ to predict the vector field and train it using the $L_{pflow}$ loss:
\begin{equation}
     \min_\theta \mathbb E_{t,p_1( x_1 \mid c), p_0(x_0)}\left\|v_p(x, t|c;\theta)-(x_1 - x_0)\right\|^2
    \label{eq:lcfp}
\end{equation}

\subsection{CFG Flow Matching Postnet}
\label{sec: fmpn}
During the first stage, the mel-spectrogram decoder primarily leverages simple losses (e.g., L1 or L2) to reconstruct the generated mel-spectrograms. Following FastSpeech2 \citep{ren2020fastspeech}, we combine pitch and technique features as inputs and employ stacked FFT (Feed Forward Transformer) blocks with L2 loss for generation training:
\begin{equation}
     L_{mel} = \left\|mel_p - mel_g\right\|^2,
    \label{eq:fs}
\end{equation}

However, the generator optimized under the assumption of an unimodal distribution yields mel-spectrograms that lack naturalness and diversity. To further enhance the quality and expressiveness of the mel-spectrograms, we adopt the CFG flow matching mel postnet (CFGFMP). In this work, we utilize the coarsely generated mel-spectrograms $mel_p$ and the combined pitch and technique features $E_m$ as conditioning information $c$ to guide the training and generation of optimized mel-spectrograms $mel_g$. The $L_{mflow}$ loss is analogous to the $L_{pflow}$ loss, as shown in equation \ref{eq:lcfp}.

For the reverse process, we randomly sample noise and use the Euler solver to generate samples.
To further control the quality of the generated singing voice and its alignment with the intended technique, we implement the classifier-free guidance (CFG) strategy. Specifically, we introduce an unconditional label $2$ alongside the conditional labels $\{0, 1\}$. During the first two stages, we randomly drop the technique labels for entire phrases or partial phonemes at a rate of $0.1$. During sampling, we modify the vector field as follows:
\begin{equation}
    v_{\mathrm{CFG}}(x, t|c;\theta) = \gamma v_m(x, t|c;\theta) + (1-\gamma) v_m(x, t|\varnothing ;\theta) ,
\end{equation}
where $\gamma$ is the classifier free guidance scale. Additionally, since the technique detector output contains errors, this random drop approach ensures the generative model doesn't blindly trust the labels, to enhance the robustness of the model. For the pseudo-code of the algorithm, please refer to Algorithm \ref{alg: train} and Algorithm \ref{alg: infer} provided in Appendix \ref{sec: appendix2pc}.

\subsection{Technique Predictor}
For controllable singing synthesis, such as timbre and emotion, many approaches use deterministic labels or corresponding audio to control the generation \citep{liu2022diffsinger, zhang2024stylesinger}. We use natural language as a more intuitive and convenient means to control singing techniques. 

However, open-source datasets don't provide corresponding prompts for each sample. Therefore, we devise a method to generate descriptions. Unlike Prompt-Singer \citep{wang2024prompt}, which focuses on simple controls like gender, vocal range, and volume, we need to control the singing techniques. We incorporate the singer's identity (e.g., Alto, Tenor), singing techniques, and language into prompt statements to annotate each sample. First, we collect the singer identity information and the global technique labels from the dataset. Then, we use GPT-4o to generate synonyms for each singer's identity and singing technique. We create over 60 prompt templates, each containing placeholders for the song's global technique label, language, and identity. We randomly select these templates and fill in the corresponding synonyms of techniques, identities, and languages to form prompt descriptions for each item. We provide the prompt templates and keywords in the appendix \ref{sec: appendix1pt}.

As shown in Figure~\ref{fig: arch}(c), our technique predictor comprises two components: a frozen natural language encoder for extracting semantic features and a technique decoder. For the natural language encoder, we evaluate both BERT \citep{devlin2018bert} and FLAN-T5 \citep{Chung2022ScalingIL} encoders. For the technique decoder, we inject semantic conditions through cross-attention transformers, allowing the model to integrate linguistic cues more effectively. 
Finally, several classification heads are added to perform multi-task, multi-label classification for different techniques.
Singing techniques are classified into three categories: mixed-falsetto and intensity, and four binary categories: breathy, bubble, vibrato, and pharyngeal. The glissando technique can be identified from the music score by determining if a word corresponds to multiple notes.The $L_{\text{tech}}$ classification loss is:
\begin{equation}
\begin{aligned}
&L_{\text{CE}}^{(i)} = -\sum_{k=1}^{3} y_k^{(i)} \log(p_k^{(i)}) \\
&L_{\text{BCE}}^{(j)} = -\left[ y^{(j)} \log(p^{(j)}) + (1 - y^{(j)}) \log(1 - p^{(j)}) \right]
\end{aligned}
\end{equation}
\begin{equation}
\begin{aligned}
&L_{\text{tech}} = \sum_{i=1}^{2} L_{\text{CE}}^{(i)} + \sum_{j=1}^{4} L_{\text{BCE}}^{(j)}
\end{aligned}
\end{equation}
where $L_{\text{CE}}^{(i)}$ represents the cross-entropy loss for the $i$-th three-class technique group, and $L_{\text{BCE}}^{(j)}$ represents the binary cross entropy loss for the $j$-th binary technique group.

\subsection{Technique Detector}

Due to the scarcity of technique-labeled singing voice synthesis datasets and the cost and complexity of annotating, we train a singing technique detector to obtain phone-level technique labels. We can also annotate the glissando technique sequence by the same rule as the technique predictor.

\begin{figure}[t]
\centering
\includegraphics[width=0.45\textwidth, trim={100mm 35mm 80mm 25mm}, clip]{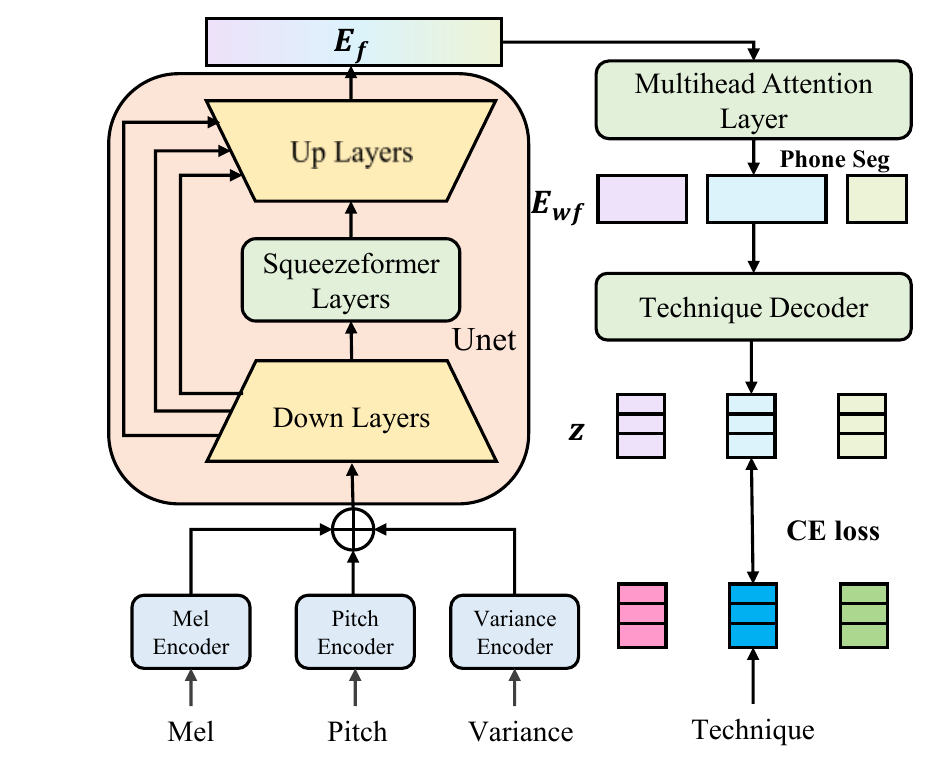}
\caption{The architecture of the technique detector.}
\label{fig: detector}
\end{figure}

As shown in Figure~\ref{fig: detector}, we start by extracting features from the audio, including the mel-spectrogram, fundamental frequency (F0), and other variances features (e.g., energy, and breathiness). These features are encoded and combined as the input feature. We then pass them through a U-Net architecture to extract frame-level intermediate features. To capture the high-level audio features, we utilize the Squeezeformer \citep{Kim2022SqueezeformerAE} network, one of the most popular ASR models.
Inspired by ROSVOT \citep{li2024robust}, rather than just using simple averaging or median operations to obtain phoneme-level audio features, we employ a weight prediction average approach. 
Suppose the frame-level output features are $E_f \in \mathbb{R}^{T \times C}$, where $T$ is the number of frames and $C$ is the number of channels. We predict weights $W_f = \sigma(E_fW_{\text{A}})$ using a linear layer and the sigmoid operation, where $W_{\text{A}} \in \mathbb{R}^{C \times N}$, $N$ is the number of heads, and $W_f \in \mathbb{R}^{T \times N}$. We then apply the weights to element-wise multiply $E_f$ to obtain weighted features $E_{wf} = E_f \odot W_f$.
Assume that phone \(i\) corresponds to a sequence starting from frame \(j\) with a length of \(k\).
we perform a weighted average method across the frame-level embeddings to obtain the final phoneme-level features $E_{wp}$:
\begin{equation}
E_{wp}^i = \frac{\sum_{t=1}^{k} E_{wf}^{i+j+t}}{\sum_{t=1}^{k} W_f^{i+j+t}}
\end{equation}
where $E_{wp} \in \mathbb{R}^{L \times C \times N} $, $L$ is the length of phones.
Next, we average different heads to get the final phoneme-level features $z \in \mathbb{R}^{L \times C} $. 
Finally, we also use cross-entropy (CE) loss $L_p$ to optimize the multi-task, multi-label technique classification task like the technique predictor.

\subsection{Training and Inference Procedures}

The training process of TechSinger comprises two stages. 
During the first stage, we optimize the entire model, excluding the post-processing flow-matching network, and use gradient descent to minimize the $L_1$ loss:
\begin{equation}
L_1 = L_{pflow} + L_{mel} + L_{dur}
\end{equation}
where $L_{pflow}$, $L_{mel}$, and $L_{dur}$ represent the F0 flow matching, mel-spectrogram, and duration losses, respectively.
During the second stage, we freeze the components trained in the first phase and optimize the classifier-free flow matching postnet ($L_{mflow}$) using adding feature $E_m$ of the predicted fundamental frequency, coarse mel-spectrogram, and technique encoding as the condition.
During the inference generation process, we can get the technique sequence based on input or prompt statements, which are then combined with lyrics and notes to generate a coarse mel-spectrogram. Subsequently, the flow-matching network refines this coarse mel-spectrogram to produce the final output.

\section{Experiments}

\subsection{Experimental Setup}
\begin{table*}[t]
\centering
\small
\vspace{2mm}
\scalebox{0.98}{
\begin{tabular}{l|cc|cc}
\toprule
\rule{0pt}{10pt} \bfseries{Method}& \bfseries{MOS-Q} $\uparrow$ & \bfseries{MOS-C} $\uparrow$  & \bfseries{FFE} $\downarrow$ & \bfseries{MCD} $\downarrow$  \\
\midrule  
Refernece & 4.54 $\pm$ 0.05 & - & - & -
\\
Reference (vocoder) & 4.15 $\pm$ 0.06 & 4.30 $\pm$ 0.09 & 0.034 & 0.919 \\
\midrule  
DiffSinger & 3.59 $\pm$ 0.07 & 3.84 $\pm$ 0.08 & 0.255 & 3.897
\\
VISinger2 & 3.52 $\pm$ 0.05 & 3.85 $\pm$ 0.11 & 0.296 & 3.944
\\
StyleSinger & 3.69 $\pm$ 0.09 & 3.93 $\pm$ 0.08 & 0.328 & 3.981 
\\
\midrule  
\bf TechSinger (ours) & \bf 3.89 $\pm$ 0.07 & \bf 4.10 $\pm$ 0.08 & \bf 0.245 & \bf 3.823
\\
\bottomrule      
\end{tabular}}
\caption{
Technique controllable singing voice synthesis performance comparison with different systems. We employ MOS-Q and MOS-C for subjective measurement and use FFE and MCD for objective measurement.
}
\label{tab: base}
\end{table*}

\begin{figure*}[t]
\centering
\includegraphics[width=0.9\textwidth, trim={3mm 90mm 0mm 62mm}, clip]{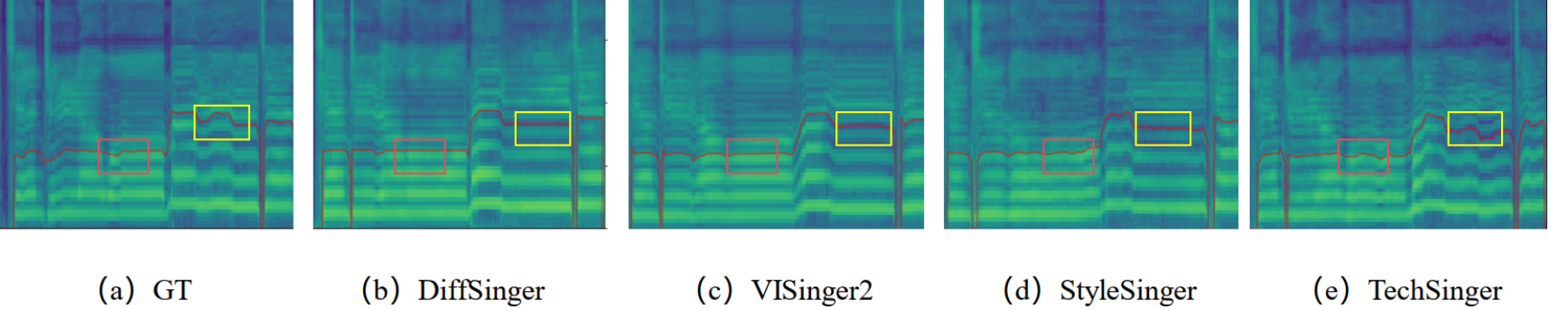}
\caption{Visualization of the mel-spectrograms and pitch contour of the ground-truth and results of different SVS systems.
}
\label{fig: dsvs}
\end{figure*}

\begin{figure*}[t]
\centering
\includegraphics[width=0.9\textwidth, trim={5mm 95mm 5mm 50mm}, clip]{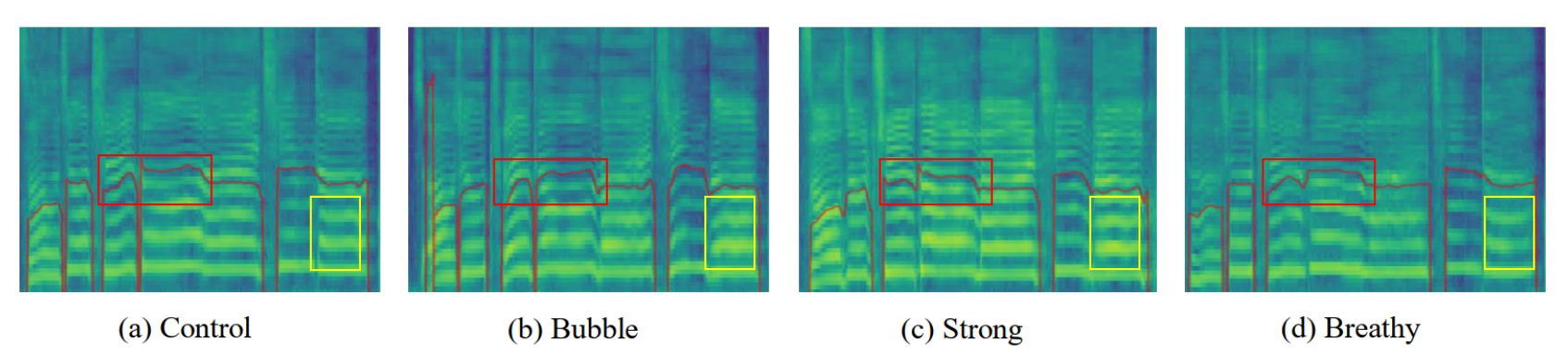}
\caption{Visualization of the mel-spectrogram results generated by TechSinger under different techniques. The red box contains the fundamental pitch, and the yellow box contains the details of harmonics.}
\label{fig: techdiff}
\end{figure*}
\subsubsection{Dataset and Process}
\label{sec: data}
Current singing synthesis datasets typically lack the diverse and detailed technique labels necessary for training high-quality models. We use the GTSinger dataset \citep{zhang2024gtsinger}, focusing on its Chinese, English, Spanish, German, and French subsets. Additionally, we collect and annotate a 30-hour Chinese dataset with two singers and four technique annotations (e.g., intensity, mixed-falsetto, breathy, bubble) at the phone and sentence levels.
Additionally, to further expand the dataset, we use a trained technique predictor and glissando judgment rule to annotate the M4Singer dataset at the phoneme level, which is used under the CC BY-NC-SA 4.0 license. Finally, we randomly select 804 segments covering different singers and techniques as a test set. The audio used for training has a sample rate of 48 kHz, with a window size of 1024, a hop size of 256, and 80 mel bins for the extracted mel-spectrograms. Chinese lyrics are phonemicized with pypinyin, English lyrics follow the ARPA standard, while Spanish, German, and French lyrics are phonemicized according to the Montreal Forced Aligner (MFA) standard.
\subsubsection{Implementation Details}
\label{sec: imple}

In this experiment, the number of training steps for the F0 and Mel vector field estimator is 100 steps. Their architectures are based on non-causal WaveNet architecture \citep{Oord2016WaveNetAG}. The number of the technique detector Squeezeformer layers and the technique predictor Transformer layers are both 2. In the first stage, training is performed for 200k steps with an NVIDIA 2080 Ti GPU, and in the second stage, for 120k steps. We train the technique detector and predictor for 120k and 80k steps.
Further details are provided in the appendix \ref{sec: appendix2vfe}.

\subsubsection{Evaluation Details}
\label{sec: eval}

For technique-controllable SVS experiments, we use both subjective and objective evaluation metrics. For objective evaluation, we use F0 Frame Error (FFE) to assess the accuracy of F0 prediction and Mean Cepstral Distortion (MCD) to measure the quality of the mel-spectrograms. For subjective evaluation, we use MOS-Q to assess the quality and naturalness of the audio and MOS-C to evaluate the expressiveness of the technique control.
We use objective metrics precision, recall, F1, and accuracy to evaluate the technique predictor and the technique detector. More details are provided in the appendix \ref{sec: appendix4metr}.

\subsubsection{Baseline Models}
In this section, we compare our approach with state-of-the-art singing voice synthesis models. However, due to the limitations of current datasets, existing singing voice synthesis models are unable to control the techniques of the generation singing audio. Therefore, we augment these baseline systems with a phoneme-level technique embedding layer to enable technique control. The baseline systems we compared are as follows:
1) GT: The ground truth audio sample; 
2) GT (vocoder): The original audio is converted to mel-spectrograms and then synthesized back to audio using the HiFi-GAN vocoder; 
3) DiffSinger \citep{liu2022diffsinger}: A diffusion-based singing voice synthesis model; 
4) VISinger2 \citep{Zhang2022VISinger2H}: An end-to-end high-fidelity singing voice synthesis model; 
5) StyleSinger \citep{zhang2024stylesinger}: A style-controllable singing voice synthesis system; 
6) TechSinger: The foundational singing voice synthesis system proposed in this paper.

\subsection{Main Results}
\subsubsection{Singing Voice Synthesis}

As shown in the Table \ref{tab: base}, we can draw the following conclusions:
(1) In terms of objective metrics, our FFE and MCD values are the lowest, which demonstrates that our TechSinger, through flow matching strategies, can better model pitch and mel-spectrograms under different singing techniques.
(2) On the subjective metric MOS-Q, our TechSinger shows higher quality than other baseline models, indicating that our model generates audio with superior quality. Similarly, on the subjective metric MOS-C, our model also outperforms other models, proving that our generation model can faithfully generate corresponding singing voices based on technique conditions.
This can be observed from Figure \ref{fig: dsvs}, where the F0 generated by our model exhibits more variation and details compared to the relatively flat F0 of other models. Additionally, our mel-spectrogram is closer to the ground truth mel-spectrograms, showcasing rich details in frequency bins between adjacent harmonics and high-frequency components. The above results demonstrate that our controllable singing voice generation model surpasses other models in terms of both quality and expressiveness in controlling technique generation.

Furthermore, to examine the technique controllability of our model, we present mel-spectrograms and F0 results for the same segments under different technique conditions. As shown in Figure \ref{fig: techdiff}, Figure (a) represents the control group without any technique, and Figure (b) displays the result for the bubble, showing more pronounced changes in F0 and mel-spectrograms with a stuttering effect, effectively reflecting the "cry-like" tone. Figure (c) shows the strong intensity, which appears brighter compared to the control group, enhancing the resonance and intensity of the singing. Figure (d) is the breathy tone result, where harmonics are less distinct and there is more noise, due to the vocal cords not fully closing as air passes through them, causing the breathy sound. From the figures, it is evident that our generated mel-spectrograms can accurately understand and generate features corresponding to different techniques. More visualization results can be found in the Appendix \ref{sec: appendix4sys}
\begin{table}[t]
\centering
\small
\vspace{2mm}
\scalebox{0.98}{
\begin{tabular}{l|cc}
\toprule
\rule{0pt}{10pt} \bfseries{Method} & \bfseries{MOS-Q} $\uparrow$ & \bfseries{MOS-C} $\uparrow$ \\
\midrule  
TechSinger(GT) & 3.89 $\pm$ 0.07 & 4.10 $\pm$ 0.08 \\
TechSinger(Rand) & 3.78 $\pm$ 0.05 & 3.76 $\pm$ 0.08 \\
TechSinger(Prompt) & 3.85 $\pm$ 0.05 & 4.04 $\pm$ 0.07 \\
\bottomrule      
\end{tabular}}
\caption{
The quality and relevance to the technique controllablity via different controlling strategies.
}
\label{tab: pred}
\end{table}

\subsubsection{Technique Predictor}
\begin{table}[t]
\centering
\small
\vspace{2mm}
\scalebox{0.98}{
\begin{tabular}{l|cccc}
\toprule
\rule{0pt}{10pt} \bfseries{Method} & \bfseries{Precision} & \bfseries{Recall} & \bfseries{F1} & \bfseries{Acc} \\
\midrule  
bert-base-uncased   & 0.819 & 0.811 & 0.807 & 0.845 \\
bert-large-uncased  & 0.809 & 0.789 & 0.786 & 0.827 \\
flan-t5-small       & 0.814 & 0.808 & 0.802 & 0.837 \\
flan-t5-base        & \bf 0.828 & 0.826 & 0.817 & \bf 0.851 \\
flan-t5-large       & 0.825 & \bf 0.836 & \bf 0.818 & 0.846 \\
\bottomrule      
\end{tabular}}
\caption{
Objective metrics for different text representations, including precision, recall, F1-score, and accuracy.
}
\label{tab: predictor}
\end{table}

We employ different text encoders to encode prompts, incorporating their embeddings into the technique sequence prediction through a cross-attention mechanism, with the results shown in Table \ref{tab: predictor}. Overall, the FLAN-T5 model's performance tends to improve with the increasing size of the encoder. The choice of encoder also has an impact, with FLAN-T5 generally outperforming BERT. Based on these observations, we select the FLAN-T5-Large model for the subsequent experiments. More results can be found in the Appendix \ref{sec: appendix1dp}

To validate the effectiveness of the technique predictor, we compare several different methods of providing techniques for generating results. Among them, TechSinger (GT) represents the results obtained from the annotated technique sequences, TechSinger (Prompt) represents the results predicted by our predictor based on prompts, and TechSinger (Random) represents the results when no techniques are provided and the model generates them automatically. From Table \ref{tab: pred}, we can see that the mean opinion scores for quality (MOS-Q) and mean opinion scores for controllability (MOS-C) indicate that the "Prompt" strategy significantly outperforms the "Random" results and are very close to the "GT" effect. This demonstrates that our singing voice synthesis model can achieve controllable technique generation through the natural language. Additionally, we can manually adjust the predicted sequences to control the technique used in the generation of singing voices further.

\subsection{Ablation Study}
\label{sec: ablation}
\subsubsection{Technique Detector}
\begin{table}[t]
\centering
\small
\vspace{2mm}
\begin{tabular}{l|cccc}
\toprule
\bfseries{Setting} & \bfseries{Precision} $\uparrow$ & \bfseries{Recall} $\uparrow$ & \bfseries{F1} $\uparrow$ & \bfseries{Acc} $\uparrow$ \\
\midrule
whole        & \bf 0.815  & \bf 0.761  & \bf 0.770 & \bf 0.833 \\
\midrule
ConvUnet  & 0.759  & 0.726  & 0.742 & 0.783 \\
Average    & 0.807  & 0.756  & 0.763  & 0.831 \\
\bottomrule   
\end{tabular}
\caption{
Ablation experiments for the technique detector. 
}
\label{tab: detect}
\end{table}
As shown in Table \ref{tab: detect}, we conduct ablation experiments on the methods used in our technique detector to prove their effectiveness. We evaluate the results using objective metrics—precision, recall, F1 score, and accuracy—on six techniques other than glissando, which can be determined by rule-based judgment. By comparing these, we find that the whole technique detector achieves the highest scores across all metrics. Specifically, we replace the Squeezeformer structure with convolution and the multi-head weight prediction method with averaging, conducting separate experiments for each. From the table, we can see that the full skill detector outperforms in all metrics, with an F1 score improvement of 0.5\% over convolution and 2.8\% over averaging, thus validating the effectiveness of the Squeezeformer and the multi-head weight prediction. For more detailed objective metric results of the individual techniques, please refer to Appendix \ref{sec: appendix3}.

\subsubsection{Singing Voice Synthesis}

\begin{table}[t]
\centering
\small
\vspace{2mm}
\begin{tabular}{l|cc|c}
\toprule
\bfseries{Setting} & \bfseries{CMOSQ} $\uparrow$ & \bfseries{CMOSC} $\uparrow$ & \bfseries{FFE} $\downarrow$\\
\midrule
TechSinger & 0.00  & 0.00  & 0.2448 \\
\midrule
w/o Pitch  & -0.25 & -0.23  & 0.2537 \\
w/o Postnet  & -0.33  & -0.27  & 0.2680 \\
w/o CFG  &  -0.10 & -0.18  & 0.2453 \\
\bottomrule   
\end{tabular}
\caption{
Ablation experiments for technique controllable singing voice synthesis with different settings.
}
\label{tab: abl}
\end{table}

As depicted in Table \ref{tab: abl}, in this experiment, we compare the results using CMOSQ, CMOSC, and FFE. As shown in the first two rows of the table, when we remove the flow-matching pitch predictor, both the F0 prediction accuracy and the quality of the generated audio decline, making it difficult to control the techniques effectively. Comparing the first and third rows, we observe a noticeable decrease in the quality of the synthesized singing when the postnet is omitted. By contrasting the first and fourth rows, we demonstrate that the classifier-free guidance strategy enhances the quality of the generated singing.

\section{Conclusion}

In this paper, we introduce TechSinger, the first multi-lingual, multi-technique controllable singing synthesis system built upon the flow-matching framework. We train a technique detector to effectively annotate and expand the dataset. To model the fundamental frequencies with high precision, we develop a Flow Matching Pitch Predictor (FMPP), which captures the nuances of diverse vocal techniques. Additionally, we employ Classifier-free Guidance Flow Matching Mel Postnet (CFGFMP) to refine the coarse mel-spectrograms into fine-grained representations, leading to more technique-controllable and expressive singing voice synthesis. Moreover, we train a prompt-based technique predictor to enable more intuitive interaction for controlling the singing techniques during synthesis. Extensive experiments demonstrate that our model can generate high-quality, expressive, and technique-controllable singing voices.

\section{Ethical Statement}
TechSinger's ability to synthesize singing voices with controllable techniques raises concerns about potential unfair competition and the possible displacement of professional singers in the music industry. Furthermore, its application in the entertainment sector, including short videos and other multimedia content, could lead to copyright issues. To address these concerns, we will implement restrictions on our code and models to prevent unauthorized use, ensuring that TechSinger is deployed ethically and responsibly.

\section{Acknowledgments}
This work was supported in part by the National Natural Science Foundation of China under Grant No.62222211 and Grant No.U24A20326.
\bibliography{aaai25}

\begin{thebibliography}{40}
\providecommand{\natexlab}[1]{#1}

\bibitem[{Brown et~al.(2020)Brown, Mann, Ryder, Subbiah, Kaplan, Dhariwal, Neelakantan, Shyam, Sastry, Askell et~al.}]{brown2020language}
Brown, T.; Mann, B.; Ryder, N.; Subbiah, M.; Kaplan, J.~D.; Dhariwal, P.; Neelakantan, A.; Shyam, P.; Sastry, G.; Askell, A.; et~al. 2020.
\newblock Language models are few-shot learners.
\newblock \emph{Advances in neural information processing systems}, 33: 1877--1901.

\bibitem[{Chung et~al.(2022)Chung, Hou, Longpre, Zoph, Tay, Fedus, Li, Wang, Dehghani, Brahma, Webson, Gu, Dai, Suzgun, Chen, Chowdhery, Valter, Narang, Mishra, Yu, Zhao, Huang, Dai, Yu, Petrov, hsin Chi, Dean, Devlin, Roberts, Zhou, Le, and Wei}]{Chung2022ScalingIL}
Chung, H.~W.; Hou, L.; Longpre, S.; Zoph, B.; Tay, Y.; Fedus, W.; Li, E.; Wang, X.; Dehghani, M.; Brahma, S.; Webson, A.; Gu, S.~S.; Dai, Z.; Suzgun, M.; Chen, X.; Chowdhery, A.; Valter, D.; Narang, S.; Mishra, G.; Yu, A.~W.; Zhao, V.; Huang, Y.; Dai, A.~M.; Yu, H.; Petrov, S.; hsin Chi, E.~H.; Dean, J.; Devlin, J.; Roberts, A.; Zhou, D.; Le, Q.~V.; and Wei, J. 2022.
\newblock Scaling Instruction-Finetuned Language Models.
\newblock \emph{ArXiv}, abs/2210.11416.

\bibitem[{Devlin et~al.(2018)Devlin, Chang, Lee, and Toutanova}]{devlin2018bert}
Devlin, J.; Chang, M.-W.; Lee, K.; and Toutanova, K. 2018.
\newblock BERT: Pre-training of Deep Bidirectional Transformers for Language Understanding.
\newblock \emph{arXiv preprint arXiv:1810.04805}.

\bibitem[{Guo et~al.(2024)Guo, Du, Ma, Chen, and Yu}]{guo2024voiceflow}
Guo, Y.; Du, C.; Ma, Z.; Chen, X.; and Yu, K. 2024.
\newblock VoiceFlow: Efficient Text-to-Speech with Rectified Flow Matching.
\newblock In \emph{ICASSP 2024-2024 IEEE International Conference on Acoustics, Speech and Signal Processing (ICASSP)}, 11121--11125. IEEE.

\bibitem[{Guo et~al.(2023)Guo, Leng, Wu, Zhao, and Tan}]{guo2023prompttts}
Guo, Z.; Leng, Y.; Wu, Y.; Zhao, S.; and Tan, X. 2023.
\newblock PromptTTS: Controllable text-to-speech with text descriptions.
\newblock In \emph{ICASSP 2023-2023 IEEE International Conference on Acoustics, Speech and Signal Processing (ICASSP)}, 1--5. IEEE.

\bibitem[{Hong et~al.(2023)Hong, Cui, Huang, Zhang, Liu, He, and Zhao}]{hong2023unisinger}
Hong, Z.; Cui, C.; Huang, R.; Zhang, L.; Liu, J.; He, J.; and Zhao, Z. 2023.
\newblock Unisinger: Unified end-to-end singing voice synthesis with cross-modality information matching.
\newblock In \emph{Proceedings of the 31st ACM International Conference on Multimedia}, 7569--7579.

\bibitem[{Huang et~al.(2022)Huang, Cui, Chen, Ren, Liu, Zhao, Huai, and Wang}]{huang2022singgan}
Huang, R.; Cui, C.; Chen, F.; Ren, Y.; Liu, J.; Zhao, Z.; Huai, B.; and Wang, Z. 2022.
\newblock Singgan: Generative adversarial network for high-fidelity singing voice generation.
\newblock In \emph{Proceedings of the 30th ACM International Conference on Multimedia}, 2525--2535.

\bibitem[{Ikemiya, Itoyama, and Okuno(2014)}]{ikemiya2014transferring}
Ikemiya, Y.; Itoyama, K.; and Okuno, H.~G. 2014.
\newblock Transferring vocal expression of f0 contour using singing voice synthesizer.
\newblock In \emph{Modern Advances in Applied Intelligence: 27th International Conference on Industrial Engineering and Other Applications of Applied Intelligent Systems, IEA/AIE 2014, Kaohsiung, Taiwan, June 3-6, 2014, Proceedings, Part II 27}, 250--259. Springer.

\bibitem[{Kim, Kong, and Son(2021)}]{kim2021conditional}
Kim, J.; Kong, J.; and Son, J. 2021.
\newblock Conditional variational autoencoder with adversarial learning for end-to-end text-to-speech.
\newblock In \emph{International Conference on Machine Learning}, 5530--5540. PMLR.

\bibitem[{Kim et~al.(2022)Kim, Gholami, Shaw, Lee, Mangalam, Malik, Mahoney, and Keutzer}]{Kim2022SqueezeformerAE}
Kim, S.; Gholami, A.; Shaw, A.~E.; Lee, N.; Mangalam, K.; Malik, J.; Mahoney, M.~W.; and Keutzer, K. 2022.
\newblock Squeezeformer: An Efficient Transformer for Automatic Speech Recognition.
\newblock \emph{ArXiv}, abs/2206.00888.

\bibitem[{Kim et~al.(2023)Kim, Kim, Jun, and Kim}]{kim2023muse}
Kim, S.; Kim, Y.; Jun, J.; and Kim, I. 2023.
\newblock MuSE-SVS: Multi-Singer Emotional Singing Voice Synthesizer that Controls Emotional Intensity.
\newblock \emph{IEEE/ACM Transactions on Audio, Speech, and Language Processing}.

\bibitem[{Kong, Kim, and Bae(2020)}]{kong2020hifi}
Kong, J.; Kim, J.; and Bae, J. 2020.
\newblock Hifi-gan: Generative adversarial networks for efficient and high fidelity speech synthesis.
\newblock \emph{Advances in neural information processing systems}, 33: 17022--17033.

\bibitem[{Kreuk et~al.(2022)Kreuk, Synnaeve, Polyak, Singer, D{\'e}fossez, Copet, Parikh, Taigman, and Adi}]{kreuk2022audiogen}
Kreuk, F.; Synnaeve, G.; Polyak, A.; Singer, U.; D{\'e}fossez, A.; Copet, J.; Parikh, D.; Taigman, Y.; and Adi, Y. 2022.
\newblock Audiogen: Textually guided audio generation.
\newblock \emph{arXiv preprint arXiv:2209.15352}.

\bibitem[{Kumar et~al.(2021)Kumar, Goel, Narang, and Lall}]{kumar2021normalization}
Kumar, N.; Goel, S.; Narang, A.; and Lall, B. 2021.
\newblock Normalization Driven Zero-Shot Multi-Speaker Speech Synthesis.
\newblock In \emph{Interspeech}, 1354--1358.

\bibitem[{Le et~al.(2024)Le, Vyas, Shi, Karrer, Sari, Moritz, Williamson, Manohar, Adi, Mahadeokar et~al.}]{le2024voicebox}
Le, M.; Vyas, A.; Shi, B.; Karrer, B.; Sari, L.; Moritz, R.; Williamson, M.; Manohar, V.; Adi, Y.; Mahadeokar, J.; et~al. 2024.
\newblock Voicebox: Text-guided multilingual universal speech generation at scale.
\newblock \emph{Advances in neural information processing systems}, 36.

\bibitem[{Li et~al.(2024)Li, Zhang, Wang, Hong, Huang, and Zhao}]{li2024robust}
Li, R.; Zhang, Y.; Wang, Y.; Hong, Z.; Huang, R.; and Zhao, Z. 2024.
\newblock Robust Singing Voice Transcription Serves Synthesis.
\newblock arXiv:2405.09940.

\bibitem[{Lipman et~al.(2022)Lipman, Chen, Ben-Hamu, Nickel, and Le}]{lipman2022flow}
Lipman, Y.; Chen, R.~T.; Ben-Hamu, H.; Nickel, M.; and Le, M. 2022.
\newblock Flow Matching for Generative Modeling.
\newblock In \emph{The Eleventh International Conference on Learning Representations}.

\bibitem[{Liu et~al.(2022)Liu, Li, Ren, Chen, and Zhao}]{liu2022diffsinger}
Liu, J.; Li, C.; Ren, Y.; Chen, F.; and Zhao, Z. 2022.
\newblock Diffsinger: Singing voice synthesis via shallow diffusion mechanism.
\newblock In \emph{Proceedings of the AAAI conference on artificial intelligence}, volume~36, 11020--11028.

\bibitem[{Liu et~al.(2021)Liu, Wen, Lu, Song, and Sung}]{liu2021vibrato}
Liu, R.; Wen, X.; Lu, C.; Song, L.; and Sung, J.~S. 2021.
\newblock Vibrato learning in multi-singer singing voice synthesis.
\newblock In \emph{2021 IEEE Automatic Speech Recognition and Understanding Workshop (ASRU)}, 773--779. IEEE.

\bibitem[{Liu, Gong et~al.(2022)}]{liu2022flow}
Liu, X.; Gong, C.; et~al. 2022.
\newblock Flow Straight and Fast: Learning to Generate and Transfer Data with Rectified Flow.
\newblock In \emph{The Eleventh International Conference on Learning Representations}.

\bibitem[{Lu et~al.(2020)Lu, Wu, Luan, Tan, and Zhou}]{lu2020xiaoicesing}
Lu, P.; Wu, J.; Luan, J.; Tan, X.; and Zhou, L. 2020.
\newblock Xiaoicesing: A high-quality and integrated singing voice synthesis system.
\newblock \emph{arXiv preprint arXiv:2006.06261}.

\bibitem[{Mehta et~al.(2024)Mehta, Tu, Beskow, Sz{\'e}kely, and Henter}]{mehta2024matcha}
Mehta, S.; Tu, R.; Beskow, J.; Sz{\'e}kely, {\'E}.; and Henter, G.~E. 2024.
\newblock Matcha-TTS: A fast TTS architecture with conditional flow matching.
\newblock In \emph{ICASSP 2024-2024 IEEE International Conference on Acoustics, Speech and Signal Processing (ICASSP)}, 11341--11345. IEEE.

\bibitem[{Ramesh et~al.(2021)Ramesh, Pavlov, Goh, Gray, Voss, Radford, Chen, and Sutskever}]{ramesh2021zero}
Ramesh, A.; Pavlov, M.; Goh, G.; Gray, S.; Voss, C.; Radford, A.; Chen, M.; and Sutskever, I. 2021.
\newblock Zero-shot text-to-image generation.
\newblock In \emph{International Conference on Machine Learning}, 8821--8831. PMLR.

\bibitem[{Ren et~al.(2020{\natexlab{a}})Ren, Hu, Tan, Qin, Zhao, Zhao, and Liu}]{ren2020fastspeech}
Ren, Y.; Hu, C.; Tan, X.; Qin, T.; Zhao, S.; Zhao, Z.; and Liu, T.-Y. 2020{\natexlab{a}}.
\newblock Fastspeech 2: Fast and high-quality end-to-end text to speech.
\newblock \emph{arXiv preprint arXiv:2006.04558}.

\bibitem[{Ren et~al.(2020{\natexlab{b}})Ren, Tan, Qin, Luan, Zhao, and Liu}]{ren2020deepsinger}
Ren, Y.; Tan, X.; Qin, T.; Luan, J.; Zhao, Z.; and Liu, T.-Y. 2020{\natexlab{b}}.
\newblock Deepsinger: Singing voice synthesis with data mined from the web.
\newblock In \emph{Proceedings of the 26th ACM SIGKDD International Conference on Knowledge Discovery \& Data Mining}, 1979--1989.

\bibitem[{Resna and Rajan(2023)}]{resna2023multi}
Resna, S.; and Rajan, R. 2023.
\newblock Multi-voice singing synthesis from lyrics.
\newblock \emph{Circuits, Systems, and Signal Processing}, 42(1): 307--321.

\bibitem[{Song et~al.(2022)Song, Song, Zhang, Zhang, Zeng, Liu, and Yu}]{song2022singing}
Song, Y.; Song, W.; Zhang, W.; Zhang, Z.; Zeng, D.; Liu, Z.; and Yu, Y. 2022.
\newblock Singing voice synthesis with vibrato modeling and latent energy representation.
\newblock In \emph{2022 IEEE 24th International Workshop on Multimedia Signal Processing (MMSP)}, 1--6. IEEE.

\bibitem[{van~den Oord et~al.(2016)van~den Oord, Dieleman, Zen, Simonyan, Vinyals, Graves, Kalchbrenner, Senior, and Kavukcuoglu}]{Oord2016WaveNetAG}
van~den Oord, A.; Dieleman, S.; Zen, H.; Simonyan, K.; Vinyals, O.; Graves, A.; Kalchbrenner, N.; Senior, A.~W.; and Kavukcuoglu, K. 2016.
\newblock WaveNet: A Generative Model for Raw Audio.
\newblock In \emph{Speech Synthesis Workshop}.

\bibitem[{Vyas et~al.(2023)Vyas, Shi, Le, Tjandra, Wu, Guo, Zhang, Zhang, Adkins, Ngan et~al.}]{vyas2023audiobox}
Vyas, A.; Shi, B.; Le, M.; Tjandra, A.; Wu, Y.-C.; Guo, B.; Zhang, J.; Zhang, X.; Adkins, R.; Ngan, W.; et~al. 2023.
\newblock Audiobox: Unified audio generation with natural language prompts.
\newblock \emph{arXiv preprint arXiv:2312.15821}.

\bibitem[{Wang et~al.(2024)Wang, Hu, Huang, Hong, Li, Liu, You, Jin, and Zhao}]{wang2024prompt}
Wang, Y.; Hu, R.; Huang, R.; Hong, Z.; Li, R.; Liu, W.; You, F.; Jin, T.; and Zhao, Z. 2024.
\newblock Prompt-Singer: Controllable Singing-Voice-Synthesis with Natural Language Prompt.
\newblock \emph{arXiv preprint arXiv:2403.11780}.

\bibitem[{Wang et~al.(2022)Wang, Wang, Zhu, Wu, Li, Xue, Zhang, Xie, and Bi}]{wang2022opencpop}
Wang, Y.; Wang, X.; Zhu, P.; Wu, J.; Li, H.; Xue, H.; Zhang, Y.; Xie, L.; and Bi, M. 2022.
\newblock Opencpop: A high-quality open source chinese popular song corpus for singing voice synthesis.
\newblock \emph{arXiv preprint arXiv:2201.07429}.

\bibitem[{Wei et~al.(2023)Wei, Cao, Dan, and Chen}]{wei2023rmvpe}
Wei, H.; Cao, X.; Dan, T.; and Chen, Y. 2023.
\newblock RMVPE: A Robust Model for Vocal Pitch Estimation in Polyphonic Music.
\newblock \emph{arXiv preprint arXiv:2306.15412}.

\bibitem[{Wu and Luan(2020)}]{wu2020adversarially}
Wu, J.; and Luan, J. 2020.
\newblock Adversarially trained multi-singer sequence-to-sequence singing synthesizer.
\newblock \emph{arXiv preprint arXiv:2006.10317}.

\bibitem[{Yang et~al.(2023)Yang, Liu, Huang, Lei, Weng, Meng, and Yu}]{yang2023instructtts}
Yang, D.; Liu, S.; Huang, R.; Lei, G.; Weng, C.; Meng, H.; and Yu, D. 2023.
\newblock Instructtts: Modelling expressive tts in discrete latent space with natural language style prompt.
\newblock \emph{arXiv preprint arXiv:2301.13662}.

\bibitem[{Zhang et~al.(2022{\natexlab{a}})Zhang, Li, Wang, Deng, Liu, Ren, He, Huang, Zhu, Chen et~al.}]{zhang2022m4singer}
Zhang, L.; Li, R.; Wang, S.; Deng, L.; Liu, J.; Ren, Y.; He, J.; Huang, R.; Zhu, J.; Chen, X.; et~al. 2022{\natexlab{a}}.
\newblock M4singer: A multi-style, multi-singer and musical score provided mandarin singing corpus.
\newblock \emph{Advances in Neural Information Processing Systems}, 35: 6914--6926.

\bibitem[{Zhang et~al.(2022{\natexlab{b}})Zhang, Cong, Xue, Xie, Zhu, and Bi}]{zhang2022visinger}
Zhang, Y.; Cong, J.; Xue, H.; Xie, L.; Zhu, P.; and Bi, M. 2022{\natexlab{b}}.
\newblock Visinger: Variational inference with adversarial learning for end-to-end singing voice synthesis.
\newblock In \emph{ICASSP 2022-2022 IEEE International Conference on Acoustics, Speech and Signal Processing (ICASSP)}, 7237--7241. IEEE.

\bibitem[{Zhang et~al.(2024{\natexlab{a}})Zhang, Huang, Li, He, Xia, Chen, Duan, Huai, and Zhao}]{zhang2024stylesinger}
Zhang, Y.; Huang, R.; Li, R.; He, J.; Xia, Y.; Chen, F.; Duan, X.; Huai, B.; and Zhao, Z. 2024{\natexlab{a}}.
\newblock StyleSinger: Style Transfer for Out-of-Domain Singing Voice Synthesis.
\newblock In \emph{Proceedings of the AAAI Conference on Artificial Intelligence}, volume~38, 19597--19605.

\bibitem[{Zhang et~al.(2024{\natexlab{b}})Zhang, Jiang, Li, Pan, He, Huang, Wang, and Zhao}]{zhang2024tcsinger}
Zhang, Y.; Jiang, Z.; Li, R.; Pan, C.; He, J.; Huang, R.; Wang, C.; and Zhao, Z. 2024{\natexlab{b}}.
\newblock TCSinger: Zero-Shot Singing Voice Synthesis with Style Transfer and Multi-Level Style Control.
\newblock In \emph{Proceedings of the 2024 Conference on Empirical Methods in Natural Language Processing}, 1960--1975.

\bibitem[{Zhang et~al.(2024{\natexlab{c}})Zhang, Pan, Guo, Li, Zhu, Wang, Xu, Lu, Hong, Wang et~al.}]{zhang2024gtsinger}
Zhang, Y.; Pan, C.; Guo, W.; Li, R.; Zhu, Z.; Wang, J.; Xu, W.; Lu, J.; Hong, Z.; Wang, C.; et~al. 2024{\natexlab{c}}.
\newblock Gtsinger: A global multi-technique singing corpus with realistic music scores for all singing tasks.
\newblock \emph{arXiv preprint arXiv:2409.13832}.

\bibitem[{Zhang et~al.(2022{\natexlab{c}})Zhang, Xue, Li, Xie, Guo, Zhang, and Gong}]{Zhang2022VISinger2H}
Zhang, Y.; Xue, H.; Li, H.; Xie, L.; Guo, T.; Zhang, R.; and Gong, C. 2022{\natexlab{c}}.
\newblock VISinger 2: High-Fidelity End-to-End Singing Voice Synthesis Enhanced by Digital Signal Processing Synthesizer.
\newblock \emph{ArXiv}, abs/2211.02903.

\end{thebibliography}

\clearpage

\appendix
\setcounter{secnumdepth}{2}

\section{Technique Predictor}
\label{sec: appendix1}
\subsection{Prompt Templates}
\label{sec: appendix1pt}
\begin{table}[ht]
    \small
    \centering
    \begin{tabular}{l|l}
    \toprule
    Keyword & Synonym  \\ 
    \midrule
    \multicolumn{2}{l}{\textbf{Identity}} \\
    \midrule
    alto & contralto, low lady voice, female low range \\
    \midrule
    tenor & high male voice, tenor vocalist, male high range \\
    \midrule
    \multicolumn{2}{l}{\textbf{Technique}} \\
    \midrule
    breathy & airy, whispery, soft-spoken \\
    \midrule
    strong &  powerful, robust, forceful \\
    \midrule
    falsetto &  head voice, light voice, false voice \\
    \midrule
    \multicolumn{2}{l}{\textbf{Language}} \\
    \midrule
    \multicolumn{2}{l}{English, Chinese, French, Spanish, German} \\
    \midrule
    \multicolumn{2}{l}{\textbf{Templates}} \\
    \midrule
    \multicolumn{2}{l}{Could you generate a song where the singer employs [tech]?} \\
    \midrule
    \multicolumn{2}{l}{Compose a melody featuring the [tech] style of singing.} \\
    \midrule
    \multicolumn{2}{l}{Design a vocal performance using [tech], delivered by a [id].} \\
    \midrule
    \multicolumn{2}{l}{Create a [lan] song that integrates [tech] into the vocal style.} \\
    \midrule
    \multicolumn{2}{l}{Develop a [lan] song with [tech] in the vocals of a [id].} \\
    \bottomrule 
    \end{tabular}
    \caption{The keyword and synonyms for each prompt attribute and different templates.}
    \label{tab:prompt_keys}
\end{table}

As shown in Table \ref{tab:prompt_keys}, we provide some prompt labels and their synonyms. We also give different template samples. Prompt templates contain the technique attribute and may randomly include language and singer identity.
\subsection{Details of the Predictor Results}
\label{sec: appendix1dp}
\begin{table*}[ht]
\centering
\small
\scalebox{0.98}{
\begin{tabular}{l|c|ccccccc}
\toprule
\multirow{2}{*}{\bfseries{Text Encoder}} & \multirow{2}{*}{\bfseries{Metric}} & \multicolumn{6}{c}{\textbf{Technique Prediction Accuracy}}\\
& & {breathy} &{bubble} &{pharyngeal} & {vibrato} & {mixed-falsetto} & {strong-weak}\\
\midrule
\multirow{4}{*}{\textbf{bert-base-uncased}} 
& Precision & 0.959 & 0.569 & 0.946 & 0.546 & 0.778 & 0.999 \\
& Recall    & 0.911 & 0.442 & 0.985 & 0.532 & 0.778 & 0.999\\
& F1        & 0.913 & 0.427 & 0.956 & 0.498 & 0.778 & 0.999\\
& Accuracy  & 0.892 & 0.775 & 0.937 & 0.853 & 0.778 & 0.999\\
\midrule
\multirow{4}{*}{\textbf{flan-t5-large}} 
& Precision & 0.931 & 0.535 & 0.950 & 0.506 & 0.802 & 0.999 \\
& Recall    & 0.865 & 0.575 & 0.946 & 0.674 & 0.802 & 0.999 \\
& F1        & 0.876 & 0.466 & 0.933 & 0.515 & 0.802 & 0.999 \\
& Accuracy  & 0.848 & 0.774 & 0.913 & 0.798 & 0.802 & 0.999 \\
\midrule
\multirow{2}{*}{\bfseries{Setting}} & \multirow{2}{*}{\bfseries{Metric}} & \multicolumn{6}{c}{\textbf{Technique Detection Accuracy}}\\
& & {breathy} &{bubble} &{pharyngeal} & {vibrato} & {mixed-falsetto} & {strong-weak}\\
\midrule
\multirow{4}{*}{\textbf{Technique Detector}} 
& Precision & 0.928 & 0.883 & 0.893 & 0.589 & 0.771 & 0.872 \\
& Recall    & 0.855 & 0.702 & 0.892 & 0.316 & 0.771 & 0.872 \\
& F1        & 0.854 & 0.757 & 0.872 & 0.374 & 0.771 & 0.872 \\
& Accuracy  & 0.851 & 0.918 & 0.848 & 0.847 & 0.771 & 0.872 \\
\bottomrule
\end{tabular}}
\caption{
Precision, recall, F1, and accuracy of the technique predictor results in different natural language text encoders and the technique detection model results.
}
\label{tab: tr2}
\end{table*}
As shown in Table \ref{tab: tr2}, we use torchmetrics to calculate precision, recall, F1, and accuracy metrics. Our prediction model can predict singing techniques such as breathy, pharyngeal, mixed-falsetto, and strong-weak with reasonable accuracy. However, we notice that the prediction of bubble and vibrato is relatively poor. In specific audio samples, we can observe that the use of bubble sounds has a high degree of randomness and is difficult to model. 

\section{Details of Postnet}
\label{sec: appendix2}

\subsection{Pseudo-Code of the Mel Postnet}
\label{sec: appendix2pc}
The algorithm of the Post-Net training and inference stage is illustrated in Algorithm \ref{alg: train} and Algorithm \ref{alg: infer}.
\renewcommand{\algorithmicrequire}{\textbf{Input:}}
\renewcommand{\algorithmicensure}{\textbf{Output:}}
\begin{algorithm}[ht]
    \caption{Pseudo-Code of the Postnet Training Stage}
    \label{alg: train}
    \begin{algorithmic}[1]
        \REQUIRE {$\bm{x}_1$: the sample mel-spectrogram, $\bm{c}$: the condition of the coarse mel-spectrogram, timbre and technique, $\bm{up}$: probability to drop the technique condition by setting the technique label to 2, $\bm{\varnothing}$: the condition of dropping the technique condition, $\bm{v}_m$: the vector field estimator.}
        \ENSURE {The neural network weights $\bm{\theta}$.}
        \STATE \textbf{function TrainStep}($\bm{v}_m, \bm{x}_0, \bm{x}_1, \bm{c}$)
            \STATE \hspace{1em} Sample $t \sim \operatorname{Uniform}[0,1]$
            \STATE \hspace{1em} Sample $\bm{x}_t = t\bm{x}_1 + (1-t)\bm{x}_0$
            \STATE \hspace{1em} $\mathcal{L}_{\text{CFM}} \leftarrow \left\|\bm{v}_m(\bm{x}_t|\bm{c};\bm{\theta})-(\bm{x}_1 - \bm{x}_0)\right\|^2$
            \STATE \hspace{1em} Gradient descent on $\mathcal{L}_{\text{CFM}}$
        \STATE Initialize neural network weights $\bm{\theta}$ randomly
        \hspace{1em} \WHILE{train the CFG flow matching}
        \STATE Take batch and sample $\bm{x}_0$ from $\mathcal{N}(\bm{0},\bm{I})$
        \STATE Sample $p \sim \operatorname{Uniform}[0,1]$
        \hspace{1em} \IF{$p < \bm{up}$}
            \STATE \textbf{TrainStep}($\bm{v_m}, \bm{x}_0, \bm{x}_1, \bm{\varnothing}$)
        \hspace{1em} \ELSE
            \STATE \textbf{TrainStep}($\bm{v_m}, \bm{x}_0, \bm{x}_1, \bm{c}$)
        \hspace{1em} \ENDIF
        \ENDWHILE
\end{algorithmic}
\end{algorithm}

\begin{algorithm}
\caption{Pseudo-Code of the Postnet Inference Stage}
\label{alg: infer}
\begin{algorithmic}[1]
    \REQUIRE {$\bm{c}$: the condition of the coarse mel-spectrogram, timbre and technique, $\bm{\gamma}$: the scale of classifier free guidance, $\bm{\varnothing}$: the condition of dropping the technique, $\bm{v}_m$: the vector field estimator, $\bm{N}$: the inference steps}
    \ENSURE {The generation sample $\bm{x}_{1}$.}
    \STATE $\bm \epsilon = 1 / N$
    \STATE $\bm t = 0$
    \WHILE{$\bm t < 1$}
        \STATE \hspace{1em} $\bm{v}_{\mathrm{CFG}} \leftarrow$ 
        \STATE \hspace{2em} $\bm \gamma \bm{v}_m(\bm x, \bm t|\bm{c};\theta) + (1-\bm \gamma) \bm{v}_m(\bm x, \bm t|\bm \varnothing ;\bm \theta)$
        \STATE \hspace{1em} $\bm{x}_{t + \bm \epsilon} = \bm{x}_t + \epsilon \bm{v}_{\mathrm{CFG}}$
        \STATE \hspace{1em} $\bm t = \bm t + \bm \epsilon$
    \ENDWHILE
    \STATE return $\bm{x}_{1}$
    
\end{algorithmic}
\end{algorithm}

\begin{table}
\centering
\small
\vspace{2mm}
\begin{tabular*}{\hsize}{l|c|c}
\toprule
\multicolumn{2}{c|}{\bfseries{Hyperparameter}}                            & \bfseries{TechSinger}    \\
\midrule
\multirow{7}*{\shortstack{Phoneme\\Encoder}}            & Phoneme Embedding         & 256   \\
~                                                       & Encoder Layers            & 4     \\
~                                                       & Encoder Hidden            & 256   \\
~                                                       & Encoder Conv1D Kernel     & 9     \\
~                                                       & Encoder Conv1D Filter Size& 1024  \\
~                                                       & Encoder Attention Heads   & 2     \\
~                                                       & Encoder Dropout           & 0.1   \\
\midrule[0.2pt]
\multirow{3}*{\shortstack{Note\\Encoder}}               & Pitches Embedding         & 256   \\
~                                                       & Type Embedding            & 256   \\
~                                                       & Duration Hidden           & 256   \\
\midrule[0.2pt]
\multirow{5}*{\shortstack{Flow\\Matching\\Pitch\\Predictor}} & Conv Layers     & 12     \\
                                                        & Kernel Size               & 3      \\
                                                        & Residual Channel          & 192    \\
~                                                       & Hidden Channel            & 256    \\
~                                                       & Training Steps                & 100    \\
\midrule[0.2pt]
\multirow{6}*{\shortstack{CFG \\Flow \\Mathing \\Postnet}}          & Conv Layers           & 20     \\
~                                                       & Kernel Size               & 3      \\
~                                                       & Residual Channel          & 256    \\
~                                                        & Hidden Channel            & 256    \\
~                                                       & Training Steps                & 100      \\
~                                                       & CFG Scale $\gamma$                & 1.2      \\

\midrule[0.2pt]
\multicolumn{2}{c|}{Sample ODE Solver}                     & Euler    \\
\bottomrule
\end{tabular*}
\caption{
Hyper-parameters of TechSinger modules.
}
\label{tab: vecarch}
\end{table}

\subsection{Vector Field Estimator}
\label{sec: appendix2vfe}
\begin{figure}
\centering
\includegraphics[width=0.45\textwidth, trim={85mm 35mm 120mm 40mm}, clip]{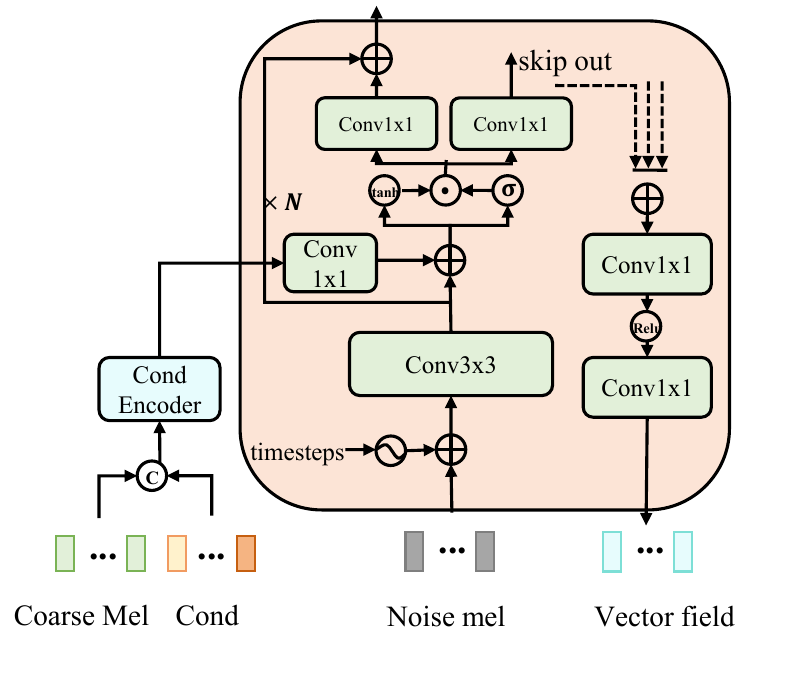}
\caption{The detailed architecture of the vector field estimator}
\label{fig: vecfield}
\end{figure}

We illustrate the architecture of the vector field estimator in Figure \ref{fig: vecfield}. 
We use the non-causal WaveNet architecture \citep{Oord2016WaveNetAG} as the backbone of our mel vector field estimator, due to its proven capability in modeling sequential data. We concatenate the mel spectrograms generated in the first stage with the generated conditioning features as conditions and use 1x1 convolutions to encode the noise mel, predicting the generated vector field. Similarly, the structure of the F0 vector field estimator is the same, except that the input changes from noise mel to noise F0, and the conditioning transforms into the extracted conditioning features. We list the hyperparameters in Table \ref{tab: vecarch}.

\section{Technique Detector}
\label{sec: appendix3}

As shown in Table \ref{tab: tr2}, we also use the precision, recall, F1, and accuracy objective metrics. The glissando can also be judged by a rule based on the number of notes corresponding to a single word, we mainly focus on detecting the other six techniques in the singing audio. From the table, we can see that the model can predict the other techniques relatively accurately. However, due to a significant imbalance between positive and negative examples in the vibrato data, we set a higher drop probability for this technique in the generative model, thereby enhancing the model's robustness.

\section{Details of Experiments}
\label{sec: appendix4}
\subsection{Dataset}
\label{sec: appendix4data}

To conduct experiments on technique-controllable singing synthesis, we have curated and annotated a Chinese high-quality, multi-technique dataset to expand both the dataset size and the variety of singing techniques. The annotated techniques include intensity, mixed-falsetto, breathy, and bubble. The intensity category is further subdivided into three labels: no technique, strong, and weak. The mixed-falsetto category includes chest voice, falsetto, and mixed voice. The remaining techniques are labeled as either present or absent.

We select one male and one female professional singer for the recordings. During the recording sessions, the singers are instructed to apply and annotate the technique labels at both the sentence and phoneme levels. Phoneme segmentation is subsequently refined using the Montreal Forced Aligner (MFA), with additional manual adjustments to ensure accuracy.
To further enrich the diversity of the dataset, we train a technique detector using the annotated data and apply it to label techniques in the open-source M4Singer dataset. The pre-processing code is available at https://github.com/gwx314/TechSinger.

\subsection{Evaluation Metrics}
\label{sec: appendix4metr}
\begin{figure*}
\centering
\includegraphics[width=0.9\textwidth]{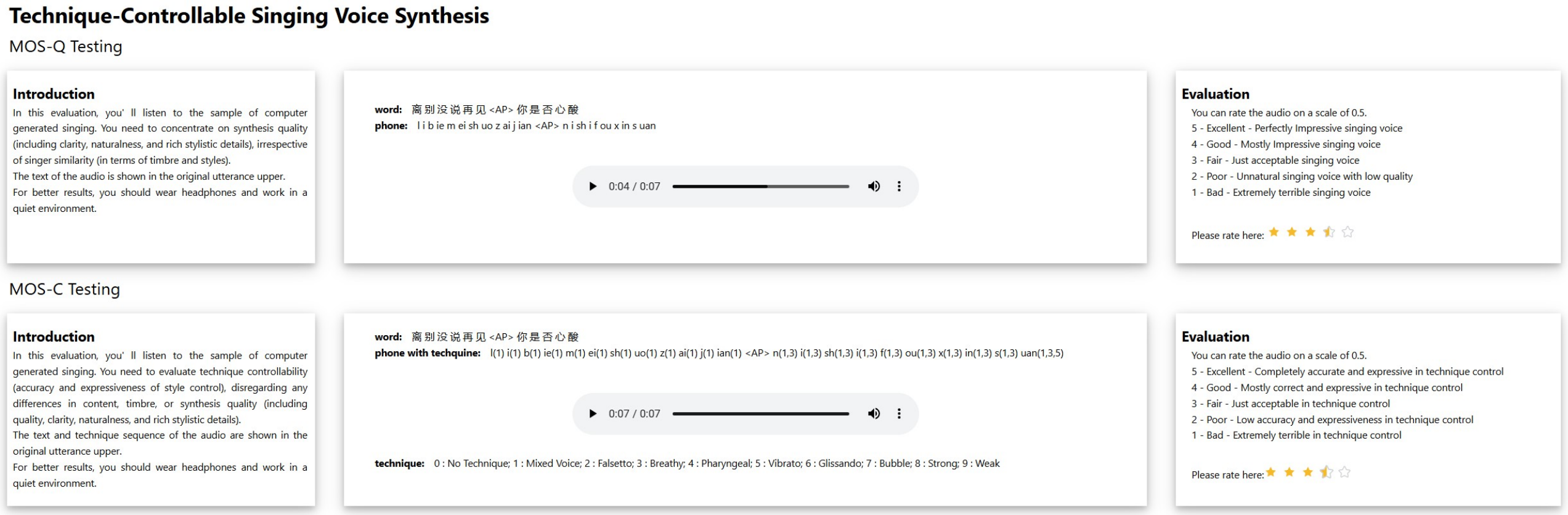}
\caption{Screenshot of MOS-Q and MOS-C testing of label guided technique-controllable SVS.}
\label{fig: sub1}
\end{figure*}
\begin{figure*}
\centering
\includegraphics[width=0.9\textwidth]{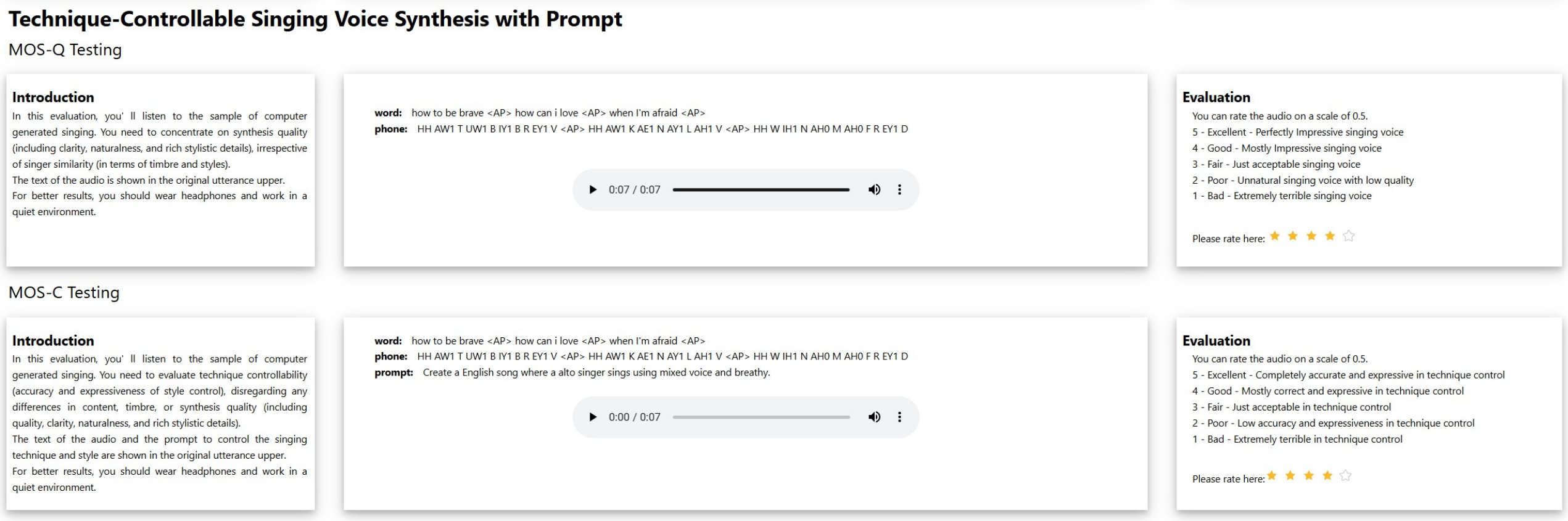}
\caption{Screenshot of MOS-Q and MOS-C testing of prompt guided technique-controllable SVS.}
\label{fig: sub2}
\end{figure*}

We randomly select 40 segments from the test set for subjective evaluation. Each generated sample and its corresponding ground-truth singing sample are evaluated by 20 professional listeners. For the MOS-Q score, listeners only evaluate the quality and expressiveness of the generated singing. For MOS-C, listeners need to compare whether the performance of the techniques in the generated singing matched the technique sequence. Both MOS-Q and MOS-C scores are rated on a five-point scale.
For the ablation study, listeners compare the differences in quality and technique expressiveness between singing samples generated with different configurations and provide CMOSQ and CMOSC scores. The screenshots of the testing instructions for listeners are shown in Figure \ref{fig: sub1} and Figure \ref{fig: sub2}. 

We use Mel Cepstral Distortion (MCD) and F0 Frame Error (FFE) as objective measures to evaluate the F0 accuracy and singing quality of the generated vocals. We calculate the Mean Cepstral Distortion (MCD) as the formula:
\begin{equation}
\begin{aligned}
&\text{MCD} = \frac{10}{\ln 10} \sqrt{2 \sum_{d=1}^{D} (m_t(d) - \hat{m}_t(d))^2},
\end{aligned}
\end{equation}
where \(m_t(d)\) and \(\hat{m}_t(d)\) is the \(d\)-th MFCC of the target and predicted frame at time \(t\), and \(D\) is the number of MFCC dimensions.
For the technique detector and technique predictor, we primarily use torchmetrics to calculate precision, recall, F1, and accuracy metrics.

\subsection{Singing Voice Synthesis}
\label{sec: appendix4sys}

\begin{figure*}
\centering
\includegraphics[width=0.9\textwidth, trim={5mm 30mm 10mm 48mm}, clip]{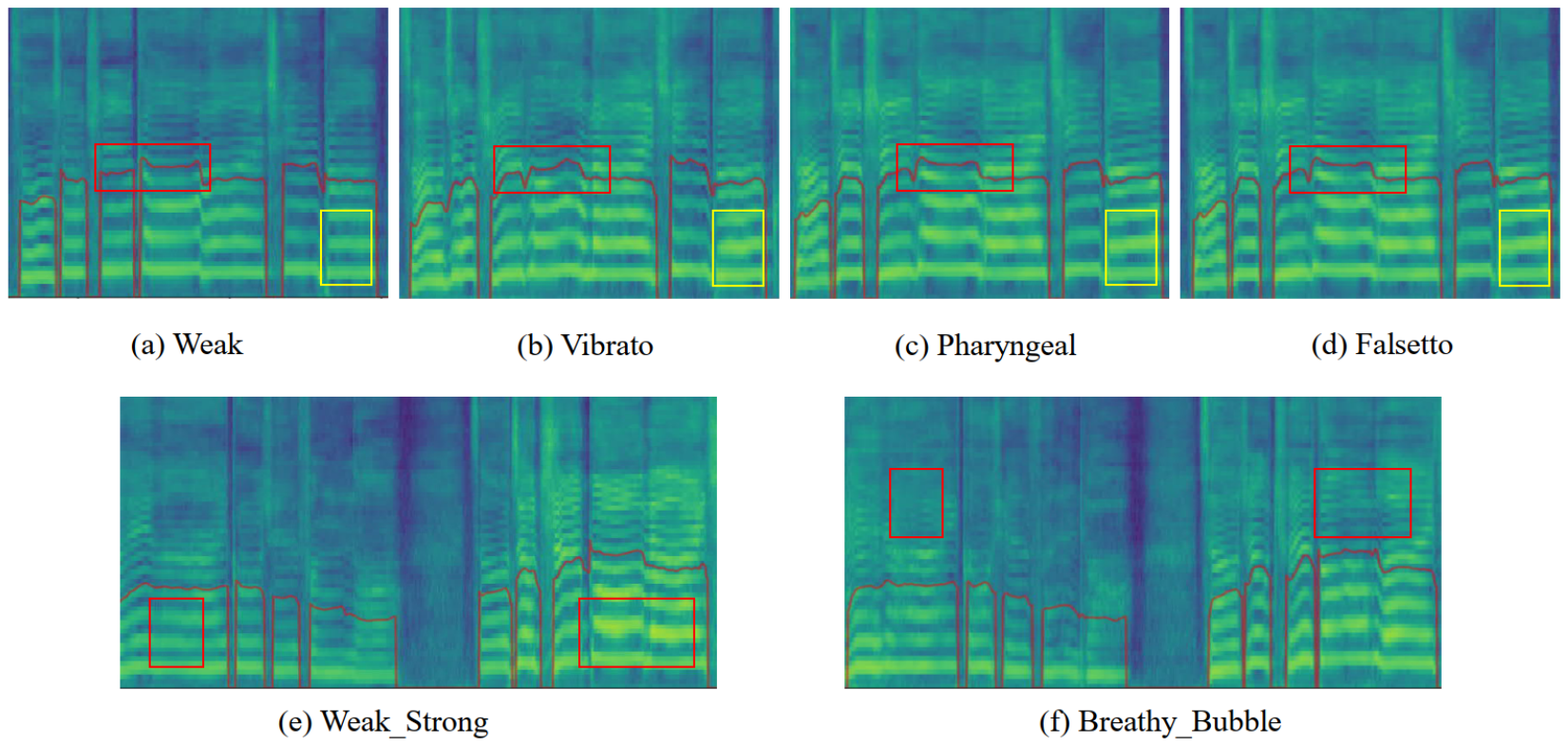}
\caption{Visualization of the mel-spectrogram results generated under different techniques.}
\label{fig: techdiffmore}
\end{figure*}

As shown in Figure \ref{fig: techdiffmore}, we present the visual results of other techniques. At the same time, we made comparisons with different techniques applied to the first and second halves of the utterances.
For Figure (e) "weak-strong", the first half represents a weak intensity while the second half represents a strong intensity, with the latter showing higher brightness in the lower frequencies.
For Figure (f) "breathy-bubble," the first half has more blurred overtones, and the second half exhibits more pitch breaks, achieving a bubbly sound effect.

\end{document}